# Sn$^{4+}$ Precursor Enables 12.4% Efficient Kesterite Solar Cell from DMSO Solution with Open Circuit Voltage Deficit Below 0.30 V

Yuancai Gong[1], Yifan Zhang[1], Erin Jedlicka[2], Rajiv Giridharagopal[2], James A. Clark[3], Weibo Yan[1], Chuanyou Niu[1], Ruichan Qiu[1], Jingjing Jiang[1], Shaotang Yu[1], Sanping Wu[1], Hugh W. Hillhouse[3], David S. Ginger[2], Wei Huang[1], and Hao Xin[1]*

**ABSTRACT. The limiting factor preventing kesterite (CZTSSe) thin film solar cell performance further improvement is the large open-circuit voltage deficit (V$_{oc,def}$) issue, which is 0.345V for the current world record device with an efficiency of 12.6%. In this work, SnCl$_4$ and SnCl$_2$·2H$_2$O are respectively used as tin precursor to investigate the V$_{oc,def}$ issue of dimethyl sulfoxide (DMSO) solution processed CZTSSe solar cells. Different complexations of tin compounds with thiourea and DMSO lead to different reaction pathways from solution to absorber material and thus dramatic difference in photovoltaic performance. The coordination of Sn$^{2+}$ with Tu leads to the formation of SnS and ZnS and Cu$_2$S in the precursor film, which converted to selenides first and then fused to CZTSSe, resulting in poor film quality and device performance. The highest efficiency obtained from this film is 8.84% with a V$_{oc,def}$ of 0.391V. The coordination of Sn$^{4+}$ with DMSO facilitates direct formation ofkesterite CZTS phase in the precursor film which directed converted to CZTSSe during selenization, resulting in compositional uniform absorber and high device performance. A device with active area efficiency 12.2% and a V$_{oc,def}$ of 0.344 V was achieved from Sn$^{4+}$ solution processed absorber. Furthermore, CZTSSe/CdS heterojunction heat treatment (JHT) significantly improved Sn$^{4+}$ device performance but had slightly negative effect on Sn$^{2+}$ device. A champion CZTSSe solar cell with a total area efficiency of 12.4% (active are efficiency 13.6%) and low V$_{oc,def}$ of 0.297 V was achieved from Sn$^{4+}$ solution. Our results demonstrate the preformed uniform kesterite phase enabled by Sn$^{4+}$ precursor is the key in achieving highly efficient kesterite absorber material. The lowest V$_{oc-def}$ and high efficiency achieved here shines new light on the future of kesterite solar cell.**

## INTRODUCTION

Kesterite Cu$_2$ZnSn(S,Se)$_4$ (CZTSSe) semiconductors have great potential as low cost photovoltaic absorber materials[1-7] because they have theoretically high efficiency to similar structured copper indium gallium selenides (Cu(In,Ga)Se$_2$, CIGS) but composed of less-toxic and earth-abundant elements. However, the efficiency of CZTSSe solar cell is only 12.6%[8, 9] whereas the efficiency of CIGS solar cell has recently reached 23.35%.[10] The barrier for CZTSSe efficiency further improvement is the large open-circuit voltage deficit (V$_{oc,def}$ = V$_{oc}$$^{SQ}$ –V$_{oc}$, V$_{oc}$$^{SQ}$ is the maximum achievable V$_{oc}$ based on Shockley-Queisser limit[11]) which is above 0.34 V for CZTS but below 0.15 V for CIGS solar cells. Possible reasons for the large V$_{oc,def}$ include potential and band gap fluctuations,[12] deep defects,[13] recombination at interfaces,[14, 15] and secondary phases.[16] Improving V$_{oc}$ is a must for practical application of kesterite solar cells.[17, 18] Recent efforts in addressing the V$_{oc,def}$ issue are mainly focused on extrinsic doping[2, 19] or alloying[3, 11, 20, 21]. However, none of the reported doping or alloying strategies has led to efficiency higher than current world record, indicating the optoelectronic properties of kesterite absorbers are more likely governed by the intrinsic defects.[6, 22]

The absorber materials of chalcogenide (CIGS and CZTSSe) thin film solar cells are fabricated from vacuum[4, 10, 23] or solution[24-28] deposited precursor films by reacting with S, H$_2$S, Se, H$_2$Se or a mixture of them (so called sulfurization and selenization) to facilitate grain growth. The property of the absorber material and the photovoltaic performance of the solar cells highly depends on the composition of the precursor film and the reaction pathway from the precursor to the absorber.[9, 21] This should be more significant for kesterite because it has two metal elements (Cu and Sn) that have variable oxidation state and Sn related deep defects have been reported to be detrimental to photovoltaic performance.[13, 29, 30] Thus, investigation on how Sn precursor oxidation state affects the reaction path from precursor film to absorber material and photovoltaic performance might be able to provide new insight of kesterite. Surprisingly, to the best of our knowledge, no such study has been performed. Compared to vacuum based approaches, which mostly use metal[9, 23] or stannous sulfide[4] as Sn precursor, solution based method provides an excellent platform for this investigation because the oxidation states of Sn can be precisely controlled by using stannous (II) or stannic (IV) salts as Sn precursor.

[1]Key Laboratory for Organic Electronics and Information Displays & Jiangsu Key Laboratory for Biosensors, Institute of Advanced Materials (IAM), Jiangsu National Synergetic Innovation Center for Advanced Materials (SICAM), Nanjing University of Posts & Telecommunications, 9 Wenyuan Road, Nanjing 210023, China.

[2]Department of Chemistry, University of Washington, Seattle, WA, 98195 USA

[3]Department of Chemical Engineering, University of Washington, Seattle, WA, 98195 USA
* Corresponding author (iamhxin@njupt.edu.cn)

Some of the authors had previously reported 4.1% efficient CZTSSe solar cells with the absorber fabricated from DMSO solution using $SnCl_2 \cdot 2H_2O$, $Cu(OAc)_2$, $ZnCl_2$, and thiourea (Tu) as precursors.[31] Later, by facilitating complete redox reaction between $SnCl_2 \cdot 2H_2O$ and $Cu(OAc)_2$ ($2Cu^{2+}+Sn^{2+}=2Cu++Sn^{4+}$) the efficiency of CZTSSe solar cell was improved to 8.3%.[25] Although the redox reaction was intentionally to reduce detrimental $Cu^{2+}$, more $Sn^{2+}$ was oxidized to $Sn^{4+}$, which may also account for the improved performance. The efficiency of DMSO solution based CZTSSe solar cell was further improved to beyond 11% by alkali metal ions along with fabrication condition optimization.[2, 19, 32] In the above DMSO solutions, Sn and Cu precursors were both in +2 oxidation states ($SnCl_2$, $CuCl_2$ or $Cu(OAc)_2$), even with complete redox reacton between $Sn^{2+}$ and $Cu^{2+}$, there was still $Sn^{2+}$ in the solution due to the requirement of a copper poor (Cu:Sn < 2) kesterite composition for better device performance. This means at least two reaction paths ($Sn^{2+}$ and $Sn^{4+}$) were involved in converting the precursors from solution to the absorber materials.

To investigate the reaction paths of $Sn^{2+}$ and $Sn^{4+}$ and their effect on kesterite photovoltaic performance, in this report, we have respectively used $SnCl_2 \cdot 2H_2O$ and $SnCl_4$ as tin source and CuCl as copper source (to avoid redox reaction) to make $Sn^{2+}$ and $Sn^{4+}$ precursor solution and fabricate CZTSSe solar cells. We found that the two solutions indeed took different reaction pathways from solution to absorber film, which resulted in dramatic difference in photovoltaic performance. A high crystalline precursor film composed of multiple phases including SnS was formed from $Sn^{2+}$ solution whereas uniform amorphous kesterite phase without secondary phases was obtained from $Sn^{4+}$ solution. Although both precursor films converted to high crystalline absorbers upon selenization, the different reaction pathway leads to dramatic difference in film morphology, composition and photovoltaic properties. A maximum active area efficiency of 12.2% with $V_{oc,def}$ of 0.344 V was obtained from $Sn^{4+}$ device, which only 8.84% with $V_{oc,def}$ of 0.391 V for $Sn^{2+}$ device. Furthermore, heterojunction heat treatment (JHT) of CZTSSe/CdS significantly improved $Sn^{4+}$ device performance but had slightly negative effect on $Sn^{2+}$ device. A champion CZTSSe solar cell with a total area efficiency of 12.4% (active area efficiency of 13.6%) and record low $V_{oc,def}$ of 0.297 V was achieved from $Sn^{4+}$ solution.

## EXPERIMENTAL SECTION
### Preparation of the molecular precursor solutions

The precursor solutions were prepared in a glovebox with controlled $O_2$ and $H_2O$ level below 5 ppm at room temperature. First, CuCl-thiourea (Tu) solution was made by adding 1.602 g Tu (99%, Aladdin, recrystallized twice) to a vial containing 4 mL dimethyl sulfoxide (DMSO, 99.8%, J&K) under stirring until completely dissolved, then 0.582g CuCl (99.999%, Alfa) was added to make a clear solution. For $Sn^{2+}$ precursor solution, 0.903g $SnCl_2 \cdot 2H_2O$ (99.99%, Aladdin), 0.567g $ZnCl_2$ (99.95%, Aladdin) and 3 mL DMSO were subsequently added to the above CuCl-Tu DMSO solution and stirred until completely dissolved. For the $Sn^{4+}$ solution, 1.042 g of $SnCl_4$ (99.99%, Alfa) was added into a second vial, the vial was sealed to prevent $SnCl_4$ evaporation. Then, 3 mL DMSO was injected into the vial by a syringe. $SnCl_4$ reacts with DMSO violently and forms white precipitate. Then 0.763 g $Zn(OAc)_2$ (99.99%, Aladdin) was added to the $SnCl_4$-DMSO suspension and stirred until a clear solution was formed. This solution was then mixed with CuCl-Tu solution to yield a pale yellow solution. We use $Zn(OAc)_2$ instead of $ZnCl_2$ as Zn source for $Sn^{4+}$ precursor solution because $SnCl_4$ coordinates with DMSO and forms $Sn(DMSO)_4Cl_4$ (Figure S6), which has low solubility in DMSO. The use of $Zn(OAc)_2$ reduces Cl- concentration and stabilize the precursor solution.

### Fabrication of CZTSSe absorber films

Molybdenum-coated soda-lime glass (MSLG) substrates were cleaned by sequential sonication in acetone and 2-propanol, each for 15 min, and dried under $N_2$ flow. The precursor film fabrication was done in a glovebox with $O_2$ and $H_2O$ level below 5 ppm. The CZTS precursor solution was filtered with 0.8 μm PTFE filter and spin-coated on the clean MSLG substrate at 1500 rpm for 60s. The wet film was immediately annealed on a hot plate at 420°C for 2 min. The coating-annealing-cooling cycle was repeated seven times to build up a precursor film with a thickness ~2 μm. The film was then put into a graphite box with Se tablets (~500 mg) and placed in Rapid Heating furnace tube for selenization. The selenization was performed at 550°C for 20 min with Ar flow rate of 20 sccm.

### Fabrication of solar cell devices

The cadmium sulfide buffer layer of thickness 40~50 nm was deposited by the chemical bath deposition (CBD) method.[27] For samples without junction heat treatment (JHT), a window layer containing 50 nm i-ZnO and 250 nm ITO was directly deposited on top of CdS by RF sputtering at room temperature. For heterojunction heat treatment, the samples (MSLG/CZTSSe/CdS) were loaded on sample holder and put into the chamber for sputtering; the chamber was pumped down to vacuum below $2*10^{-3}$ Pa. The sample holder (and samples) was heated to 200°C within 30 min and kept at 200°C for a designed time (2 hours or 20 hours), then the window layer containing 50 nm i-ZnO and 150 nm ITO was deposited at 200°C right after the JHT. Finally, top contact grids of nickel (50 nm) and aluminum (200 nm) were fabricated in a separate thermal deposition system through a shadow mask. The solar cell area was defined by mechanical scribing for approximately 0.105 cm² and the accurate area was individually measured as described in the Supporting Information. ITO sputtered at room temperature has higher resistance than that sputtered at 200°C due to poor crystallinity, so the ITO of samples without JHT is thicker (250 nm) than samples with JHT (150 nm) in order to have similar conductivity. The thinner ITO has slightly higher transparency than thicker ITO.

**Film characterization**

X-ray diffraction (XRD) spectra (2θ scan) were collected by a Siemens D5005 X-ray powder diffraction system using Cu K (λ = 1.5406 Å) X-ray as the source. The Raman spectra were acquired on Renishaw in Via microscope using 532 or 785 nm Laser diode as the excitation source. The film surface morphology and elemental composition analysis were measured on Hitachi S4800 scanning electron microscope (SEM) using 5 kV power for imaging and 15 or 20 kV power for energy dispersive X-ray (EDX) spectroscopy. The film composition profile was acquired by glow discharge optical emission spectroscopy (GDOES) using a Horiba GD-Profiler 2 instrument with an anode of 4 mm. Plasma optimization and GDOES calibration were based on information presented by Payling and Nelis[1]. Photoluminescence spectra were acquired on a modified Horiba LabRAM HR-800 using a 785 nm laser diode as excitation source and a liquid nitrogen cooled InGaAs array as detector.

**Device characterization**

The current density–voltage (J-V) curves were measured using Keithley 2400 Source Meter under simulated AM 1.5 sunlight at 100 mW/cm$^2$ irradiance generated by an AAA sun simulator (CROWNTECH, Inc.) with the intensity calibrated by an NREL calibrated Si reference cell. The external quantum efficiency (EQE) of the solar cells was measured on Enlitech QE-R3018 using calibrated Si and Ge diodes (Enli technology Co. Ltd.) as references.

**RESULTS AND DISCUSSION**

Figure 1a and Figure 1b show the active area current density-voltage (J-V) and external quantum efficiency (EQE) curves of the best performing Sn$^{2+}$ and Sn$^{4+}$ devices fabricated under standard condition with 98 nm MgF$_2$ antireflective coating (ARC). The device structure and a representative cross-section SEM image of Sn$^{4+}$ solar cell are shown in Figure 1c. Since our devices have relatively larger grids (~10.5% total area) than ideal condition (~4%) due to mask design and grid fabrication condition, we use active area efficiency to better reflect absorber property and total area efficiency for fair comparison with literature, both are specified throughout the report. The method of device total and active area calculation is given in Figure S1. A maximum active area power conversion efficiency (PCE) of 12.2% with J$_{sc}$, V$_{oc}$, FF of 37.2 mA/cm$^2$, 0.475 V and 68.8% was achieved from Sn$^{4+}$ solution. The PCE, J$_{sc}$, V$_{oc}$, FF obtained from Sn$^{2+}$ solution were respectively 8.84%, 35.5 mA/cm$^2$, 0.421 V and 59.1%. Sn$^{4+}$ solar cells not only have higher value in all device parameters than Sn$^{2+}$ solar cells but also exhibit much smaller scattering in all device parameters (Figure 1d and Table S1), revealing higher quality and uniformity of Sn$^{4+}$ absorber. The photoluminescence spectra (Fig-ure S2) show that Sn$^{4+}$ absorber exhibits higher PL intensity, higher peak energy and narrower full width at the half maximum (FWHM) than Sn$^{2+}$ film. The EQE spectra (Figure 1b) reveal Sn$^{4+}$ device has better carrier collection efficiency than Sn$^{2+}$ device, especially in the near infrared range. The bandgaps of the CZTSSe absorber materials estimated from the EQE data (Figure 1b, inset) are 1.050 eV for Sn$^{2+}$ device and 1.058 eV for Sn$^{4+}$ device. Based on V$_{oc}^{SQ}$ calculated from equation V$_{oc}^{SQ}$ = 0.932 × E$_g$ − 0.167, the V$_{oc,def}$ of Sn$^{2+}$ and Sn$^{4+}$ devices are calculated to be 0.391 V and 0.344 V, respectively. A V$_{oc,def}$ of 0.344 V is one of the lowest value achieved from kesterite without extra modification such as doping,[6, 22] grain boundary passivation,[33] or alloying,[6, 22] demonstrating the advance of Sn$^{4+}$ precursor in achieving high quality kesterite absorber.

Figure 1e shows the powder X-ray diffraction (XRD) patterns of the precursor and absorber films fabricated parallel to the best performing devices in Figure 1a. For Sn$^{2+}$ precursor film, sharp diffraction peaks that can be assigned to kesterite Cu$_2$ZnSnS$_4$ (PDF# 26-0575) and SnS (PDF#39-0354) are observed together with detectable peak from Cu$_{2-x}$S (Figure 1e). The formation of SnS and Cu$_{2-x}$S suggests the existence of ZnS because of the Zn-rich composition, although ZnS cannot be distinguished from CZTS due to diffraction peak overlap. The high crystallinity of Sn$^{2+}$ precursor film is also confirmed by SEM (Figure 1f) and Raman (Figure S3). In contrast, Sn$^{4+}$ precursor film only contains amorphous Cu$_2$ZnSnS$_4$ as revealed by XRD (Figure 1e) and Raman (Figure S3) results. Both precursor films converted to high crystalline absorber film upon selenization with different morphology (Figure 1f, and Figure S4). Sn$^{2+}$ absorber contains large grains with some small grains at the bottom whereas Sn$^{4+}$ absorber exhibits a clear double-layer structure with both layers consisting of well-packed large grains. In addition, ZnSe aggregates are observed on the surface of Sn$^{2+}$ absorber from the SEM image (Figure 1f) whereas the surface of Sn$^{4+}$ film is clean. We note that a tri-layer structure with a fine-grain middle layer is often observed for Sn$^{4+}$ films (Figure S4b), which achieves similar efficiency as the double-layer film, whereas loosely packed small grains are frequently seen for Sn$^{2+}$ films (Figure S4a). Further optimizing film fabrication condition to avoid the double-layer morphology is expected to further improve Sn$^{4+}$ device performance. From the energy dispersive X-ray (EDX) spectroscopy measurement (Table S2), both Sn$^{2+}$ and Sn$^{4+}$ absorber films have a global Cu-poor and Zn-rich composition with Cu:Zn:Sn ratios of 1:0.71:0.62 for Sn$^{2+}$ absorber and 1:0.71:0.65 for Sn$^{4+}$ absorber. Compared to the feed-in composition (Cu:Zn:Sn=1:0.68:0.65), Sn$^{2+}$ film exhibits slight Sn loss.

To understand the morphology and performance difference between Sn$^{2+}$ and Sn$^{4+}$ absorbers, solution chemistry and reaction pathways from solution to absorber film were investigated(Figures S5 and S6). Solution chemistry shows that CuCl, ZnCl$_2$ and SnCl$_2$ coordinate with Tu and form Cu(Tu)$_3$Cl, Zn(Tu)$_2$Cl$_2$ and Sn(Tu)$_2$Cl$_2$, which respectively decomposes to Cu$_{2-x}$S, ZnS and SnS upon thermal annealing (Figure

S5). However, $SnCl_4$ only coordinates with DMSO and forms $Sn(DMSO)_4Cl_4$. Annealing a wet film containing $SnCl_4$ and Tu does not form $SnS_2$ due to direct decomposition of $Sn(DMSO)_4Cl_4$ to volatile $SnCl_4$ (boiling point 114.15°C). Investigation on the grain growth from precursor film to absorber material (Figure S7) reveals the amorphous CZTS in $Sn^{4+}$ precursor film directly converted to CZTSSe by substitution reaction at the very early stage of selenization, whereas sulfides $Cu_{2-x}S$, ZnS and SnS in $Sn^{2+}$ precursor film transferred to selenides $Cu_{2-x}Se$, ZnSe, and $SnSe_2$ first before converting to CZTSSe (more detailed grain growth mechanism of the two films will be reported in another paper). Thus, the reaction paths from $Sn^{2+}$ and $Sn^{4+}$ solutions to the CZTSSe absorbers can be written as follow:

**$Sn^{2+}$ solution:**

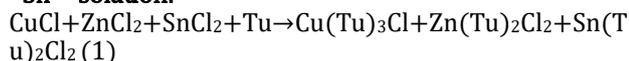

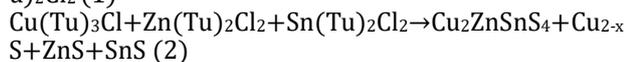

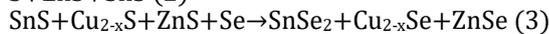

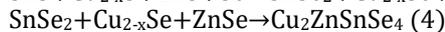

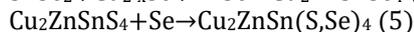

**$Sn^{4+}$ solution:**

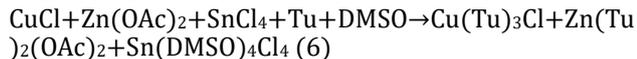

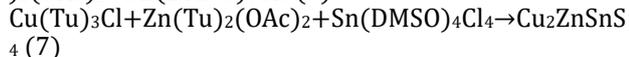

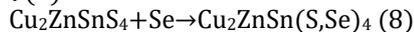

Kim et al reported that the formation of defects is directly correlated to the existence of secondary phases and deep defects can be suppressed by reducing film inhomogeneity.[34] The co-existence of all three binaries in $Sn^{2+}$ film during grain growth explains the poor uniformity and low device performance of $Sn^{2+}$ absorber. On the contrast, the direct conversion of CZTS to CZTSe (Figure S7) enables compositional and morphological uniformity of $Sn^{4+}$ absorber and thus superior device performance.

It was reported that annealing kesterite absorber at low temperature for long time can reduce Cu-Zn disorder and improve absorber electronic property.[35] Recently, Yan et al. reported that heterojunction heat treatment (JHT) can reduce interface recombination and improve kesterite solar cell efficiency.[4] We had performed JHT of CdS/CZTSSe at 200°C for 20 h in vacuum. The J-V curves of $Sn^{2+}$ and $Sn^{4+}$ devices with and without JHT are shown in **Figure 2a** and their photovoltaic parameters are summarized in Table S3. For $Sn^{2+}$ device, JHT only slightly increases $V_{oc}$ but decreases $J_{sc}$ and FF, resulting in slightly lower PCE. For $Sn^{4+}$ device, on the contrary, JHT significantly improves $V_{oc}$ and FF but slightly reduces $J_{sc}$. The best performing device with an active area efficiency of 13.0% (without ARC) was achieved from $Sn^{4+}$ absorber with $J_{sc}$ of 35.1 $mA/cm^2$, $V_{oc}$ of 0.513 V, and FF of 72.5% (Figure 2a). The $V_{oc,def}$ of this device is 0.317 V based on the bandgap of 1.070 eV extracted from the EQE (Figure S8). The statistical data of 20 h JHTed 150 devices (Figure S9 and Table S4, active area) show exactly the same trend as Figure 2a with the average $J_{sc}$, $V_{oc}$, FF and PCE of 34.7 $mA/cm^2$, 0.509 V, 70.7%, and 12.5% respec-tively for $Sn^{4+}$ devices, which again demonstrate the high reproducibility of $Sn^{4+}$ absorber.

One of the 20 h JHTed high performance device with 98 nm $MgF_2$ ARC was independently measured by NREL and a total area efficiency of 11.56% with $J_{sc}$ of 32.01 $mA/cm^2$, $V_{oc}$ of 0.520 V, and FF of 69.43% was certified on an area of 0.1066 $cm^2$ (Figure 2c and Figure S10c). The certified solar cell has an active area efficiency of 13.12% measured in house by excluding the grid area (Figure S10d). The $J_{sc}$ integrated from the NREL measured EQE (Figure 2d) is 36.61 $mA/cm^2$, corresponding to an even higher active area efficiency of 13.22%. The $V_{oc,def}$ of the certified device is 0.332 V based on the bandgap of 1.094 eV (Figure 2d), smaller than that of IBM (0.373V) and DGIST (0.345V) record devices.

From Yan et al, JHT leads to partial substitution of Zn by Cd which decreases absorber bandgap and increases device $J_{sc}$.[4] Su et al. also reported band gap decrease of kesterite with Cd alloying.[20] Glow discharge optical emission spectroscopy (GDOES, Figure 2e and 2f) shows both $Sn^{2+}$ and $Sn^{4+}$ absorber films have similar bulk compositional profiles with a large amount of Cd detected. However, the bandgap of our JHTed absorbers increases, from 1.041 to 1.065 eV for $Sn^{2+}$ absorber and 1.050 to 1.070 eV for $Sn^{4+}$ absorber, indicating different mechanism. Raman data (Figure 2g) clearly show a blue-shift of the CZTSSe peaks upon JHT, which is the direct evidence of improvement in kesterite disorder-order,[35] thus, the increase of the bandgap can be explained by the improvement in absorber order-disorder. To verify how much the order-disorder contributes to the $V_{oc}$ enhancement, annealing absorber film without CdS was conducted under same condition (at 200°C for 20 h), which only improved device $V_{oc}$ by 16 mV (Figure S11). The results indicate the order-disorder is not the vital issue for the large $V_{oc,def}$ of kesterite solar cells, in agreement with Rey's report.[36]

The similar Cd diffusion, band gap increase and $J_{sc}$ response of $Sn^{2+}$ and $Sn^{4+}$ absorbers upon JHT indicate JHT has similar effect on absorber bulk. The difference of JHT on device performance, that is, large improvement in $V_{oc}$ and FF for $Sn^{4+}$ device but slightly negative effect on $Sn^{2+}$ device, has to come from the surface property of the absorber. From the reaction pathways discussed above, the uniform composition and direct conversion of kesterite phase of $Sn^{4+}$ film ensures uniform composition and thus less defect surface. For $Sn^{2+}$ film, on the contrary, the existence of multiphase until the end of the grain growth (Figure S8) inevitably creates high concentration of defects at the surface. To further confirm this, devices with different top and bottom precursor layers (top $Sn^{4+}$/bottom $Sn^{2+}$ or top $Sn^{2+}$/bottom $Sn^{4+}$) were fabricated and their J-V characteristics are shown in Figures 2h and 2i. As expected, the $V_{oc}$ of devices with bottom $Sn^{2+}$ and top $Sn^{4+}$ layers gradually increase with the increase of $Sn^{4+}$ layers and reaches that of pure $Sn^{4+}$ device at 3 layers (Figure 2h and Table S5) whereas devices with bottom $Sn^{4+}$ and top $Sn^{2+}$ layers display linear decrease of $V_{oc}$ with the increase of $Sn^{2+}$ layers (Figure 2i and Table S6). In ad-dition, devices with $Sn^{4+}$ bottom layers have much higher FF than $Sn^{2+}$ bottom layer devices, further confirming the high quality of the $Sn^{4+}$ absorber materials. Further understand the mechanism of how JHT affects CZTSSe/CdS interface property of $Sn^{4+}$ and $Sn^{2+}$ based CZTSSe absorber is expected to provide more insight on the limiting factor of kesterite $V_{oc}$, which is under investigation and will be reported in the future.

The performance improvement of CZTS solar cell upon JHT was attributed to cation diffusion between CdS and CZTS and the JHT was conducted for a very short time.[4] We suspect the $J_{sc}$ decrease in the 20 h JHTed devices comes from the deterioration of the absorber bulk property due to long time heat treatment at the presence of Cd. To confirm this, JHT of 2 h and 20 h were carried out on both $Sn^{2+}$ and $Sn^{4+}$ films and the results are shown in Figure S12 and Table S7. As expected, 2 h JHT resulted in better performance than 20 h JHT for both $Sn^{2+}$ and $Sn^{4+}$ devices with the main gain from $J_{sc}$ for $Sn^{2+}$ device and from both $J_{sc}$ and FF for $Sn^{4+}$ device. A champion device with a total area efficiency of 12.4% (no ARC) with $J_{sc}$, $V_{oc}$, FF of 33.32 mA/cm², 0.522 V and 71.5% was achieved from $Sn^{4+}$ solution. The J-V and EQE characteristics of this device are shown in **Figure 3**. After excluding the grid area, this device has a $J_{sc}$ of 36.36 mA/cm² (Figure 3a, dashed line), in excellent agreement with the $J_{sc}$ (36.45 mA/cm²) integrated from the EQE, which corresponds to an active area efficiency of 13.6%. The band gap of this device extracted from the EQE is 1.058 eV. The $V_{oc,def}$ and $V_{oc}/V_{oc}^{SQ}$ of this device are calculated to be 0.297V and 63.7%, respectively. To our best knowledge, they are the lowest $V_{oc,def}$ and highest $V_{oc}/V_{oc}^{SQ}$ among all the reported kesterite solar cells.

## CONCLUSIONS

In conclusion, we have used $SnCl_4$ and $SnCl_2 \cdot 2H_2O$ as tin precursor to investigate the effect of tin oxidation state on DMSO solution processed kesterite absorber property and solar cell performance. We found the coordination of $Sn^{2+}$ with thiourea leads to a high crystalline precursor film containing SnS, ZnS and $Cu_{2-x}S$, whereas coordination of $Sn^{4+}$ with DMSO results in direct formation of uniform amorphous kesterite phase. Although both precursor films converted to crystalline CZTSSe absorber upon selenization, $Sn^{2+}$ film undergoes conversion of multiphase sulfides to selenides and then fusion of selenides to kesterite, resulting in poor absorber with non-uniform composition, on the contrast, direct conversion from sulfide to selenide kesterite ensures high quality absorber material of $Sn^{4+}$ film. An active area efficiency of 12.2% with $V_{oc}$ of 0.475 V was achieved from $Sn^{4+}$ device which was only 8.84% with $V_{oc}$ of 0.421 V for $Sn^{2+}$ device. Furthermore, JHT greatly improves $Sn^{4+}$ device performance but has slight negative effect on $Sn^{2+}$ device. A champion device with a total area efficiency of 12.4% (without ARC), active area efficiency of 13.6%, lowest $V_{oc,def}$ of 0.297 V, and highest $V_{oc}/V_{oc}^{SQ}$ of 63.7% has been achieved from kesterite absorber fabricated from DMSO solution using $Sn^{4+}$ precursor. Our results demonstrate a preformed uniform kesterite phase (as the $Sn^{4+}$ precursor film in this report) is critical to en-sure high quality absorber materials whereas fusion of multi secondary phases (as the $Sn^{2+}$ precursor film in this report and most cases in the literature due to the starting precursors are metal or/and secondary sulfides) creates defects due to exist of secondary phases. Our finding provides a new platform and direction for investigating the mechanism of the $V_{oc-def}$ issue of kesterite and further improving kesterite solar cell efficiency.

1    Todorov TK, Tang J, Bag S, *et al* Beyond 11% Efficiency: Characteristics of State-of-the-Art $Cu_2ZnSn(S,Se)_{(4)}$ Solar Cells. Adv Energy Mater, 2013, 3: 34-38

2    Xin H, Vorpahl SM, Collord AD, *et al* Lithium-doping inverts the nanoscale electric field at the grain boundaries in $Cu_2ZnSn(S,Se)_{(4)}$ and increases photovoltaic efficiency. Phys Chem Chem Phys, 2015, 17: 23859-23866

3    Qi Y-F, Kou D-X, Zhou W-H, *et al* Engineering of interface band bending and defects elimination via a Ag-graded active layer for efficient $(Cu,Ag)_{(2)}ZnSn(S,Se)_{(4)}$ solar cells. Energy Environ Sci, 2017, 10: 2401-2410

4    Yan C, Huang J, Sun K, *et al* $Cu_2ZnSnS_4$ solar cells with over 10% power conversion efficiency enabled by heterojunction heat treatment. Nat Energy, 2018, 3: 764-772

5    Guo L, Shi J, Yu Q, *et al* Coordination engineering of Cu-Zn-Sn-S aqueous precursor for efficient kesterite solar cells. Sci Bull, 2020, 65: 738-746

6    Giraldo S, Jehl Z, Placidi M, *et al* Progress and Perspectives of Thin Film Kesterite Photovoltaic Technology: A Critical Review. Adv Mater, 2019, 31: 1806692

7    Liu F, Wu S, Zhang Y, *et al* Advances in kesterite $Cu_2ZnSn(S, Se)_4$ solar cells. Sci Bull, 2020, 65: 698-701

8    Wang W, Winkler MT, Gunawan O, *et al* Device Characteristics of CZTSSe Thin-Film Solar Cells with 12.6% Efficiency. Adv Energy Mater, 2014, 4: 1301465

9    Son D-H, Kim S-H, Kim S-Y, *et al* Effect of solid-$H_2S$ gas reactions on CZTSSe thin film growth and photovoltaic properties of a 12.62% efficiency device. J Mater Chem A, 2019, 7: 25279-25289

10   Nakamura M, Yamaguchi K, Kimoto Y, *et al* Cd-Free $Cu(In,Ga)(Se,S)_{(2)}$ Thin-Film Solar Cell With Record Efficiency of 23.35%. IEEE J Photovolt, 2019, 9: 1863-1867

11   Collord AD, Hillhouse HW. Germanium Alloyed Kesterite Solar Cells with Low Voltage Deficits. Chem Mater, 2016, 28: 2067-2073

12   Hadke S, Levcenko S, Sai Gautam G, *et al* Suppressed Deep Traps and Bandgap Fluctuations in $Cu_2CdSnS_4$ Solar Cells with ≈8% Efficiency. Adv Energy Mater, 2019, 9: 1902509

13   Chen S, Walsh A, Gong X-G, *et al* Classification of Lattice Defects in the Kesterite Cu2ZnSnS4 and Cu2ZnSnSe4 Earth-Abundant Solar Cell Absorbers. Adv Mater, 2013, 25: 1522-1539

14   Gunawan O, Todorov TK, Mitzi DB. Loss mechanisms in hydrazine-processed $Cu_2ZnSn(Se,S)_4$ solar cells. Appl Phys Lett, 2010, 97: 346-395

15   Min X, Guo L, Yu Q, *et al* Enhancing back interfacial contact by in-situ prepared $MoO_3$ thin layer for $Cu_2ZnSnS_xSe_{4-x}$ solar cells. Sci China Mater, 2019, 62: 797-802

16   Kumar M, Dubey A, Adhikari N, *et al* Strategic review of secondary phases, defects and defect-complexes in kesterite CZTS-Se solar cells. Energy Environ Sci, 2015, 8: 3134-3159

17   Antunez PD, Bishop DM, Luo Y, *et al* Efficient kesterite solar cells with high open-circuit voltage for applications in powering distributed devices. Nat Energy, 2017, 2: 884-890

18   Siebentritt S. High voltage, please! Nat Energy, 2017, 2: 840-841

19   Haass SG, Andres C, Figi R, *et al* Complex Interplay between Absorber Composition and Alkali Doping in High-Efficiency Kesterite Solar Cells. Adv Energy Mater, 2018, 8: 1701760

20   Su Z, Tan JMR, Li X, *et al* Cation Substitution of

Solution-Processed Cu$_2$ZnSnS$_4$ Thin Film Solar Cell with over 9% Efficiency. Adv Energy Mater, 2015, 5: 1500682

21    Giraldo S, Saucedo E, Neuschitzer M, *et al* How small amounts of Ge modify the formation pathways and crystallization of kesterites. Energy Environ Sci, 2018, 11: 582-593

22    Romanyuk YE, Haass SG, Giraldo S, *et al* Doping and alloying of kesterites. J Phys Energy, 2019, 1: 044004

23    Lee YS, Gershon T, Gunawan O, *et al* Cu$_2$ZnSnSe$_4$ Thin-Film Solar Cells by Thermal Co-evaporation with 11.6% Efficiency and Improved Minority Carrier Diffusion Length. Adv Energy Mater, 2015, 5: 1401372

24    Todorov T, Hillhouse HW, Aazou S, *et al* Solution-based synthesis of kesterite thin film semiconductors. J Phys Energy, 2020, 2: 012003

25    Xin H, Katahara JK, Braly IL, *et al* 8% Efficient Cu$_2$ZnSn(S,Se)$_4$Solar Cells from Redox Equilibrated Simple Precursors in DMSO. Adv Energy Mater, 2014, 4: 1301823

26    Fu J, Fu J, Tian Q, *et al* Tuning the Se Content in Cu$_2$ZnSn(S, Se)$_4$ Absorber to Achieve 9.7% Solar Cell Efficiency from a Thiol/Amine-Based Solution Process. ACS Appl Energy Mater, 2018, 1: 594-601

27    Jiang JJ, Giridharagopal R, Jedlicka E, *et al* Highly efficient copper-rich chalcopyrite solar cells from DMF molecular solution. Nano Energy, 2020, 69: 11

28    Wu SP, Jiang JJ, Yu ST, *et al* Over 12% efficient low-bandgap CuIn(S, Se)$_{(2)}$ solar cells with the absorber processed from aqueous metal complexes solution in air. Nano Energy, 2019, 62: 818-822

29    Li J, Yuan Z-K, Chen S, *et al* Effective and Noneffective Recombination Center Defects in Cu$_2$ZnSnS$_4$: Significant Difference in Carrier Capture Cross Sections. Chem Mater, 2019, 31: 826-833

30    Kim S, Park J-S, Walsh A. Identification of Killer Defects in Kesterite Thin-Film Solar Cells. ACS Energy Lett, 2018, 3: 496-500

31    Ki W, Hillhouse HW. Earth-Abundant Element Photovoltaics Directly from Soluble Precursors with High Yield Using a Non-Toxic Solvent. Adv Energy Mater, 2011, 1: 732-735

32    Haass SG, Diethelm M, Werner M, *et al* 11.2% Efficient Solution Processed Kesterite Solar Cell with a Low Voltage Deficit. Adv Energy Mater, 2015, 5:

33    Gershon T, Shin B, Bojarczuk N, *et al* The Role of Sodium as a Surfactant and Suppressor of Non-Radiative Recombination at Internal Surfaces in Cu$_2$ZnSnS$_4$. Adv Energy Mater, 2015, 5: 1400849

34    Yang K-J, Sim J-H, Son D-H, *et al* Precursor designs for Cu$_2$ZnSn(S,Se)$_4$ thin-film solar cells. Nano Energy, 2017, 35: 52-61

35    Rey G, Redinger A, Sendler J, *et al* The band gap of Cu$_2$ZnSnSe$_4$: Effect of order-disorder. Appl Phys Lett, 2014, 105: 112106

36    Rey G, Weiss TP, Sendler J, *et al* Ordering kesterite improves solar cells: A low temperature post-deposition annealing study. Sol Energy Mater Sol Cells, 2016, 151: 131-138

**Acknowledgements** This work was supported primarily by the National Natural Science Foundation of China (NSFC, Grant No. 21571106, U1902218). Jiang J. and Yu S. acknowledge the support from Postgraduate Research and Practice Innovation Program of Jiangsu Province. Jedlicka E. and Giridharagopal R. acknowledge support of facilities in the Molecular Analysis Facility, a National Nanotechnology Coordinated Infrastructure site at the University of Washington which is supported in part by the National Science Foundation (grant NNCI-1542101), the University of Washington, the Molecular Engineering & Sciences Institute, the Clean Energy Institute, and the National Institutes of Health.

**Author contributions**    Gong Y. and Xin H. conceived the idea and co-wrote the manuscript. Xin H. and Huang W. supervised the work. Gong Y. And Zhang Y. fabricated the devices and conducted most of the measurements. Jedlicka E. and Giridharagopal R. conducted GDOES measurements and initial data process. Clark J. conducted PL measurements. Niu C., Qiu R., Jiang J., Yu S. and Wu S. provides assistance in device fabrication and measurements. Yan W., Huang W., Hillhouse H. and Ginger D. discussed the results and provided valuable suggestions to the manuscript.

**Conflict of interest**    The authors declare no conflict of interest.
**Supplementary information**    Supporting data are available in the online version of the paper.

Figures:

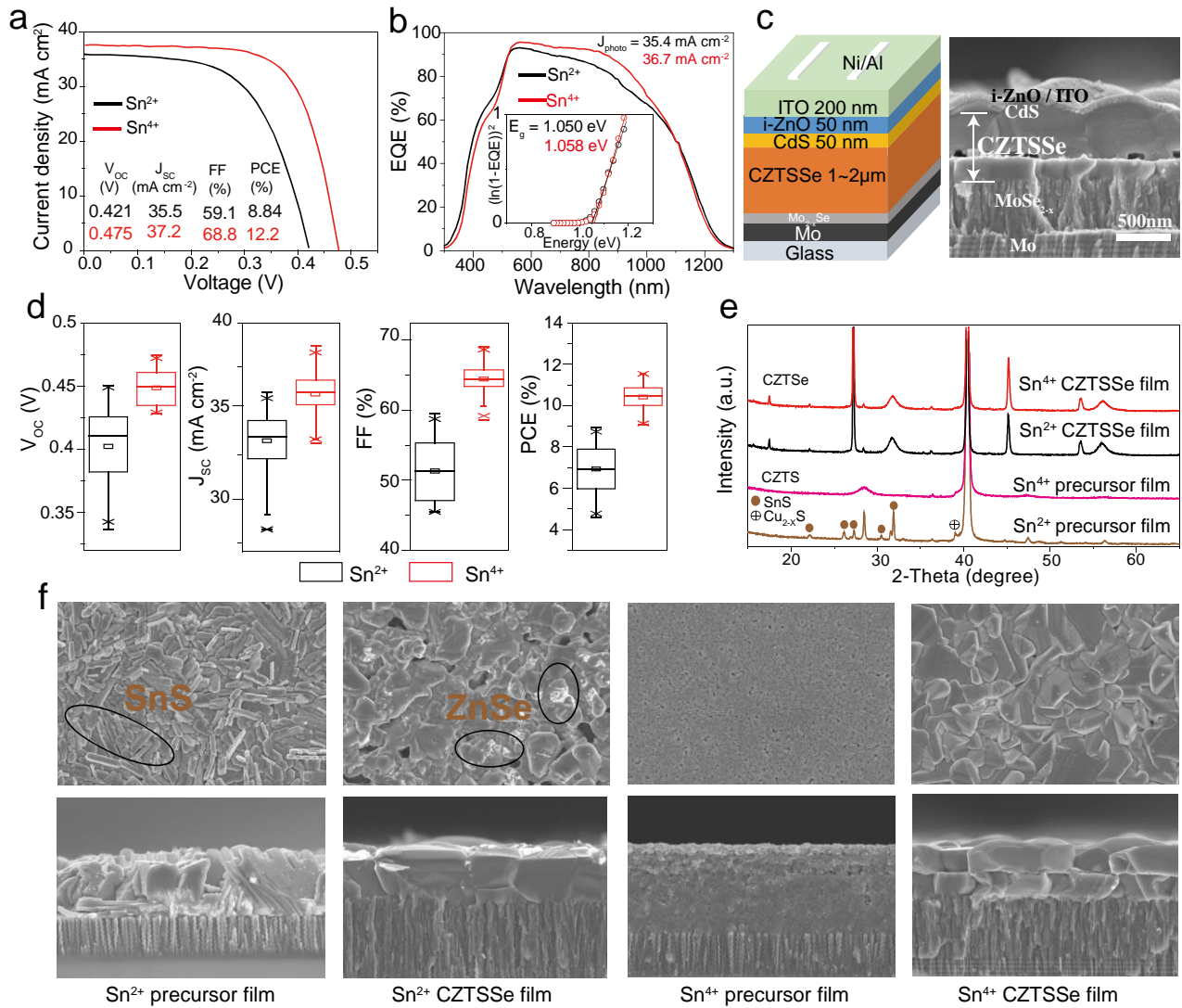

Figure 1. Characterization of CZTSSe devices and absorber films fabricated from Sn²⁺ and Sn⁴⁺ solutions. (a) J-V curves, (b) EQE spectra, (c) device structure, (d) statistical photovoltaic parameters, (e) XRD patterns, (f) plan view and cross-section SEM images of the precursor and absorber (CZTSSe) films. All images have the same scale. (b) Inset: Bandgap estimation by fitting the expected linear region of the EQE. Data in (d) are from 150 devices for each condition. Devices in (a) had 98 nm MgF₂ ARC. All efficiencies are based on active area.

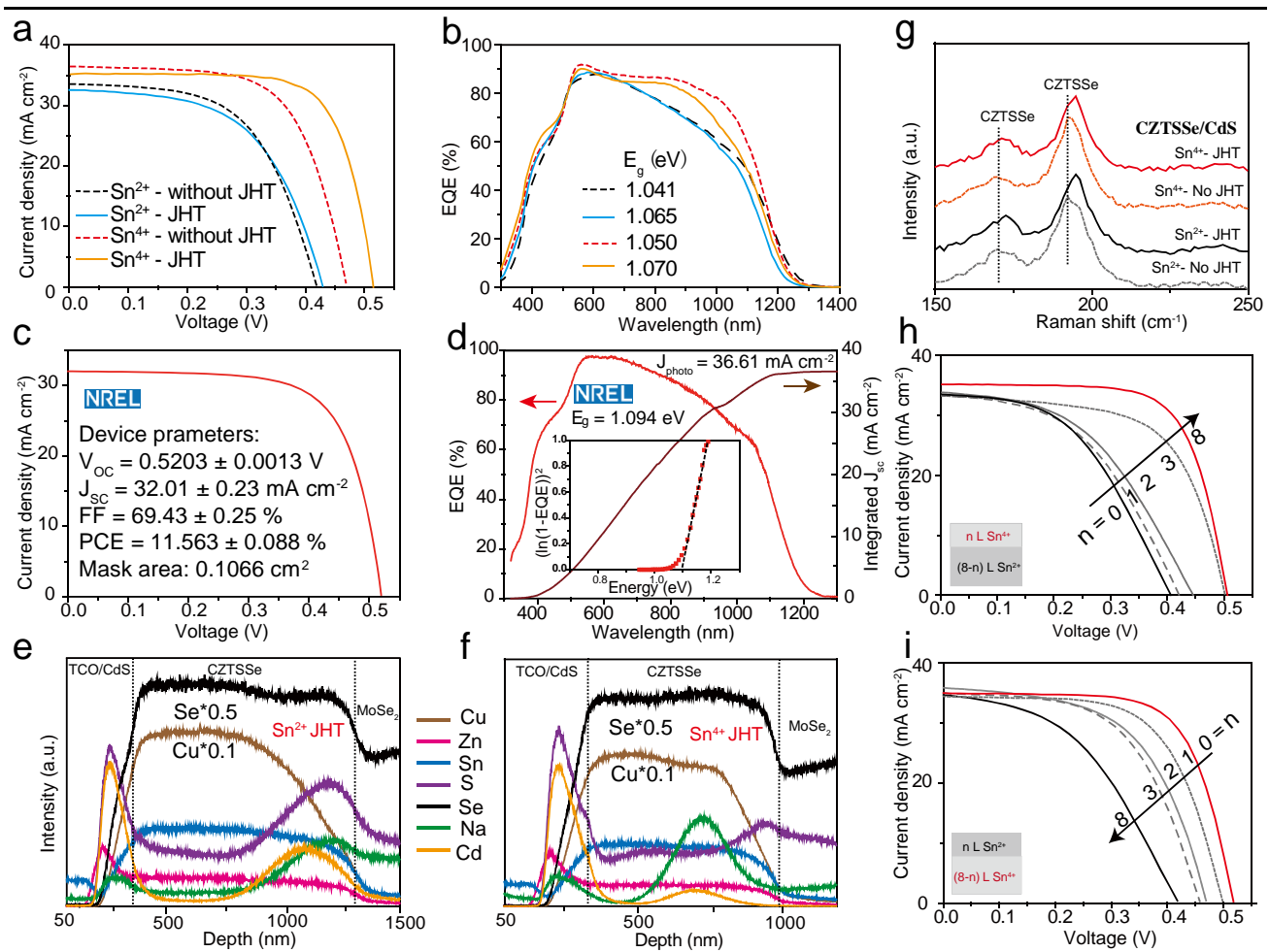

Figure 2. Effect of JHT on device performance. (a) J-V curves of Sn²⁺ and Sn⁴⁺ solar cells with and without JHT (no ARC); (b) EQE spectra of solar cells in (a); (c) The J-V and (d) the EQE of the NREL certified Sn⁴⁺ device; (e, f) Depth compositional profiles of JHTed Sn²⁺ (e) and Sn⁴⁺ (f) devices measured by GDOES; (g) Raman spectra of CdS/CZTSSe films with and without JHT (785 nm laser excitation); (h, i) J-V characteristics of devices with bottom Sn²⁺ and top Sn⁴⁺ layers (h) and bottom Sn⁴⁺ and top Sn²⁺ layers (i). The JHT was performed in vacuum for 20 h.

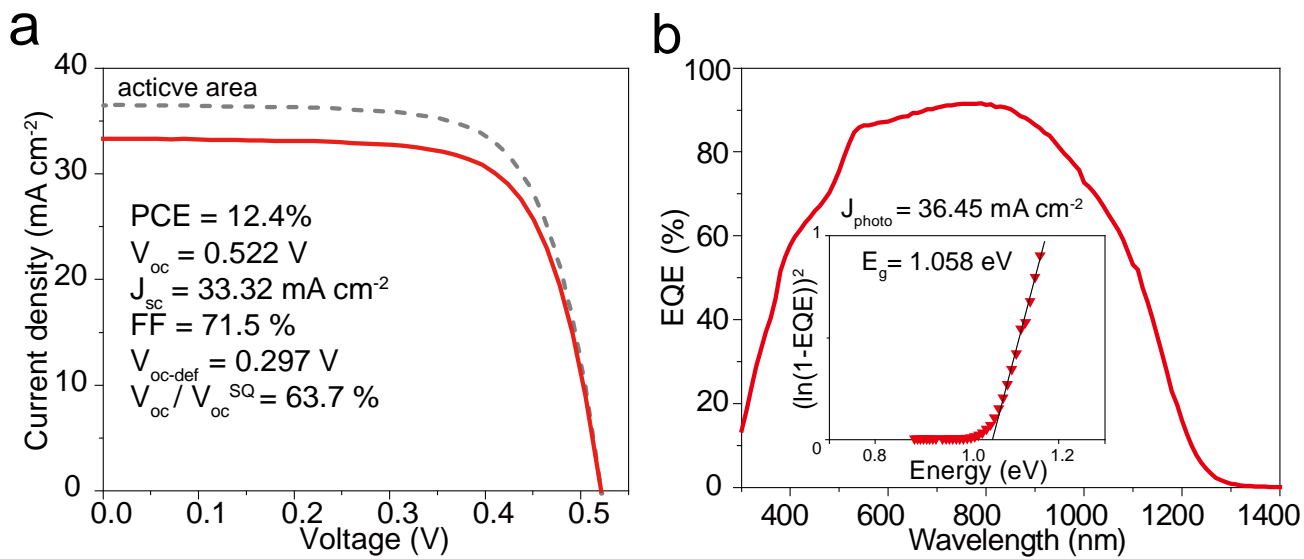

Figure 3. (a) J–V and (b) EQE of the champion CZTSSe device fabricated from $Sn^{4+}$ precursor solution with 2h JHT.

# Table of contents graphic

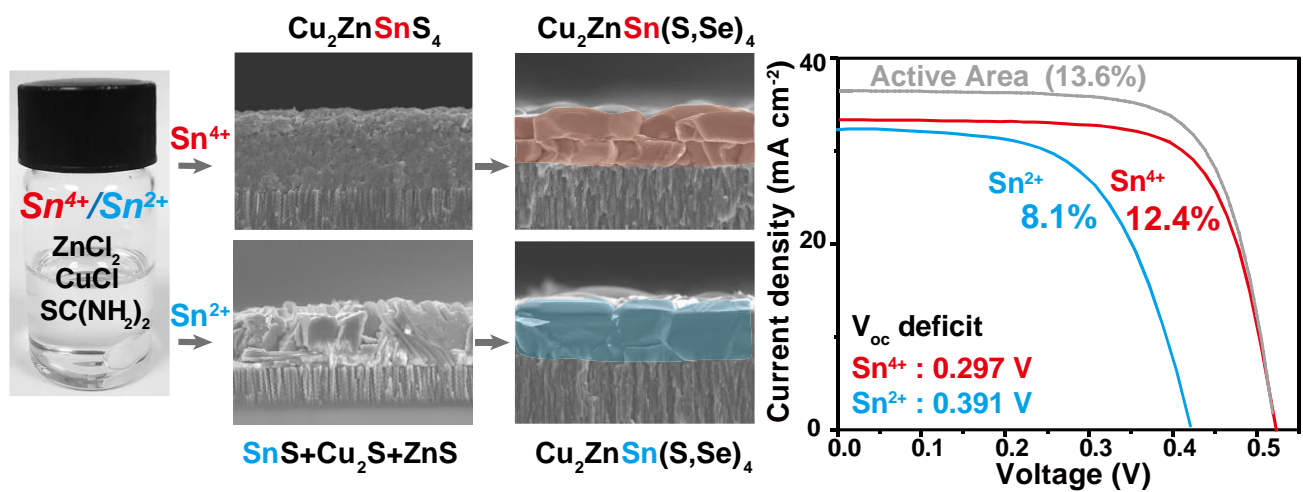